# Parallel QR Decomposition in LTE-A Systems

Sébastien Aubert*‡, Manar Mohaisen †, Fabienne Nouvel‡, and KyungHi Chang †
* ST-ERICSSON, 505, route des lucioles, CP 06560 Sophia-Antipolis, France
† Inha University, Graduate school of IT & T, 402-751 Incheon, Korea
‡ Research unit IETR, 20, avenue des buttes de Coësmes, CP 35043 Rennes, France
Email: sebastien.aubert@stericsson.com, lemanar@hotmail.com, fabienne.nouvel@insa-rennes.fr, khchang@inha.ac.kr

*Abstract*—The QR Decomposition (QRD) of communication channel matrices is a fundamental prerequisite to several detection schemes in Multiple-Input Multiple-Output (MIMO) communication systems. Herein, the main feature of the QRD is to transform the non-causal system into a causal system, where consequently efficient detection algorithms based on the Successive Interference Cancellation (SIC) or Sphere Decoder (SD) become possible. Also, QRD can be used as a light but efficient antenna selection scheme. In this paper, we address the study of the QRD methods and compare their efficiency in terms of computational complexity and error rate performance. Moreover, a particular attention is paid to the parallelism of the QRD algorithms since it reduces the latency of the matrix factorization.

## I. INTRODUCTION

Multiple-Input Multiple-Output (MIMO) technologies have gained an increasing attention in the recent standardization bodies due to their capabilities to improve the reliability of the communication link and boost the channel capacity without requiring additional spectral resources [1]. In 3GPP Long-Term Evolution (3GPP LTE) and 3GPP LTE-Advanced, MIMO techniques with up to 8 transmit antennas are combined with the Orthogonal Frequency Division Multiplexing (OFDM) to offer both low complexity equipments and robustness to the frequency selective fading [2].

The main challenge on the receiver side is to introduce low-complexity but efficient detection algorithms able to recover quasi-optimally the transmitted signals. QR Decomposition (QRD)-based Successive Interference Cancellation (SIC) schemes [3], Sphere Decoder (SD) [4] and QRD with M-algorithm (QRD-M) [5] detection algorithms use the QRD to transfer the MIMO system from a non-causal, where interference from future symbols exists, into a causal system where the detection of the current symbol only depends on the already-detected symbols. Hence, SIC becomes possible. Lenstra-Lenstra-Lovasz (LLL) [6] lattice basis reduction algorithm also requires the QRD to decrease computational complexity in the calculation of its orthogonality deficit measure.

The main idea of the QRD is to factorize the channel matrix $\mathbf{H}$ as the product of an unitary matrix $\mathbf{Q}$ and an upper triangular matrix $\mathbf{R}$. Since the matrix $\mathbf{Q}$ is unitary, the Euclidean norm, the singular values, and the determinant of $\mathbf{H}$ and $\mathbf{R}$ are equal. Thanks to the QRD, the MIMO system is transferred into a causal system by simply filtering the received vector using the Hermitian transpose of the $\mathbf{Q}$ matrix.

It turns out that the QRD, although it plays an important role, has not been sufficiently studied in the literature of the wireless communication systems.

**Contributions.** The contributions of this paper can be summarized as follows:

- First, several computation methods of the QRD are introduced. The parallelism capabilities in calculation of the QRD are emphasized;
- A reduced complexity solution is proposed and discussed;
- The complexity of the introduced algorithms is calculated in terms of real multiplications (MUL), with realistic implementation considerations.

The rest of this paper is organized as follows. In Section II, we introduce the system model and review several detection schemes that are based on the QRD. QRD techniques are introduced in details in Section III and simulation results are exposed in Section IV. Section V provides a conclusion and a discussion.

## II. SYSTEM MODEL AND MIMO DETECTION

### A. System Model

In this paper, we consider a MIMO-Spatial Multiplexing (MIMO-SM) system, where independent data symbols are transmitted via sufficiently-separated antennas. By considering $n_T$ transmit antennas and $n_R$ receive antennas - with $n_R \geq n_T$ to avoid rank-deficiency effects [7] - the received vector $\mathbf{y} \in \mathbb{C}^{n_R}$ is given by:

$$\mathbf{y} = \mathbf{Hx} + \mathbf{n}, \qquad (1)$$

where $\mathbf{x} \in \xi^{n_T}$ is the transmitted vector with $\mathrm{E}\left[\mathbf{xx}^H\right] = \mathbf{I}_{n_T}$ ($\xi$ is the modulation set and $\mathbf{I}_{n_T}$ is the $n_T \times n_T$ identity matrix) and $\mathbf{H}$ is the channel matrix with complex element $h_{i,j}$ whose real and imaginary parts are independent and follow $\mathcal{N}(0, 0.5)$. Also, $\mathbf{n}$ is the additive white Gaussian vector with $\mathrm{E}\left[\mathbf{nn}^H\right] = \sigma_n^2 \mathbf{I}_{n_R}$.

### B. MIMO Detection

In MIMO-SM, the optimum Maximum-Likelihood Detector (MLD) employs a brute-force detection to estimate the transmitted vector:

$$\hat{\mathbf{x}} = \underset{\mathbf{x} \in \xi^{n_T}}{\operatorname{argmin}} \|\mathbf{y} - \mathbf{Hx}\|^2. \qquad (2)$$

By defining the QRD $\mathbf{H} = \mathbf{QR}$ of the $n_R \times n_T$ complex channel matrix, with $\mathbf{Q}$ an unitary matrix of dimensions $n_R \times n_T$ ($\mathbf{Q}^H\mathbf{Q} = \mathbf{I}_{n_T}$) and $\mathbf{R}$ an upper triangular matrix of dimensions $n_T \times n_T$. By considering $\tilde{\mathbf{y}} = \mathbf{Q}^H\mathbf{y}$, the search problem in (2) can be re-written as:

$$\hat{\mathbf{x}} = \underset{\mathbf{x} \in \xi^{n_T}}{\operatorname{argmin}} \|\tilde{\mathbf{y}} - \mathbf{Rx}\|^2,$$

$$= \underset{\mathbf{x} \in \xi^{n_T}}{\operatorname{argmin}} \sum_{i=1}^{n_T} \left| \tilde{y}_i - R_{i,i}x_i - \sum_{j=i+1}^{n_T} R_{i,j}x_j \right|^2, \quad (3)$$

where $\tilde{y}_i$ is the $i$-th entry of $\tilde{\mathbf{y}}$ and $x_j$ is a hypothetical value for the $j$-th entry of $\mathbf{x}$ and $R_{i,j}$ is the $(i, j)$ entry of $\mathbf{R}$. Due to the QRD, the $n_T$-dimensional lattice search in (3) is transformed into $n_T$ parallel one-dimensional search problems. Therefore, instead of searching over a multi-dimensional sphere, the search is done in parallel over lines, which reduces the computational complexity.

Since embedded communication systems are limited in terms of computational complexity, several detection schemes have been proposed in the literature to solve the search problem in (3) with low computational costs.

The QRD-based SIC technique was originally introduced in [8] and further studied by Wuebben *et al.* [3], [9]. The decision process becomes, for all $i$ from $n_T$ to 1:

$$\hat{x}_i = \mathcal{Q}_\xi \left\{ \frac{\tilde{y}_i - \sum_{j=i+1}^{n_T} R_{i,j}\hat{x}_j}{R_{i,i}} \right\}, \quad (4)$$

where $\mathcal{Q}_\xi$ denotes the quantification to the modulation set $\xi$ and $\hat{x}_i$ and $\hat{x}_j$ the estimates of the $i$-th and $j$-th transmit symbols respectively, for $j > i$.

Although the SIC schemes outperform the linear schemes, they are still far from achieving the optimum performance of the MLD [10] and in particular the same diversity.

SD and in particular QRD-M detection algorithms are considered as prominent schemes that achieve quasi-ML performance while requiring lower complexity compared to the MLD. The main idea of the SD is to restrict the search problem in (3) to the lattice points inside a hyper-sphere of predefined radius $d$. The decision process takes into account the sphere constraint and becomes, for all $i$ from $n_T$ to 1:

$$\hat{\mathbf{x}} = \underset{\mathbf{x} \in \xi^{n_T}}{\operatorname{argmin}} \sum_{i=1}^{n_T} \left| \tilde{y}_i - R_{i,i}x_i - \sum_{j=i+1}^{n_T} R_{i,j}\hat{x}_j \right|^2 \leq d^2, \quad (5)$$

Although SD has a low average computational complexity, its worst-case complexity is still high. The conventional QRD-M algorithm has a fixed complexity which is a desired property for latency-limited communication systems. The idea of the QRD-M is to retain a fixed number of symbol candidates at each search level.

### III. QR-DECOMPOSITION TECHNIQUES

#### A. Gram-Schmidt Orthogonalization

Gram-Schmidt (GS) orthogonalization is frequently used in communication systems. The GS QRD algorithm consists of two steps, namely orthogonalization and normalization steps. In the orthogonalization step, a normal vector of the matrix $\mathbf{Q}$ that is already normalized in the normalization step is obtained and the remaining columns of $\mathbf{Q}$ are orthogonalized to the obtained column. Note that the matrix $\mathbf{Q}$ is initiated to $\mathbf{H}$. Therefore, the corresponding row of the matrix $\mathbf{R}$ is obtained from $\mathbf{Q}$. Algorithm 1 depicts a MATLAB-like[1] algorithmic description of the Stable GS (STGS) algorithm by introducing a minor update of the Classical GS (CLGS) algorithm.

**Algorithm 1** Stable Gram-Schmidt QRD algorithm.
1: INPUT: $\mathbf{H}$
2: OUTPUT: $\tilde{\mathbf{y}}$, $\mathbf{R}$
3: Initialization : $\mathbf{Q} \leftarrow \mathbf{H}$, $\mathbf{R} \leftarrow \mathbf{I}_{n_T}$
4: **for** $i = 1, \cdots, n_T$ **do**
5: $\quad \mathbf{R}_{i,i} \leftarrow \sqrt{\mathbf{Q}_{:,i}^H \mathbf{Q}_{:,i}}$
6: $\quad \mathbf{Q}_{:,i} \leftarrow \mathbf{Q}_{:,i}/\mathbf{R}_{i,i}$
7: $\quad$ **for** $j = i+1, \cdots, n_T$ **do**
8: $\quad\quad \mathbf{R}_{i,j} \leftarrow \mathbf{Q}_{:,i}^H \mathbf{Q}_{:,j}$
9: $\quad\quad \mathbf{Q}_{:,j} \leftarrow \mathbf{Q}_{:,j} - \mathbf{R}_{i,j}\mathbf{Q}_{:,i}$
10: $\quad$ **end for**
11: **end for**
12: $\tilde{\mathbf{y}} \leftarrow \mathbf{Q}^H\mathbf{y}$

#### B. Householder Reflections

Householder (HH) constitutes another classical QRD technique that is used to obtain the upper triangular matrix $\mathbf{R}$ from which the matrix $\mathbf{Q}$ can be obtained if required. The idea behind the HH technique is to obtain the matrix $\mathbf{R}$ using a reflection matrix. This reflection matrix, also known as Householder matrix, is used to cancel all the elements of a vector except its first element which is assigned the norm of the vector. Therefore, the columns of the matrix $\mathbf{H}$ are treated iteratively to obtain the $\mathbf{R}$ matrix. Algorithm 2 depicts a MATLAB-like algorithmic description of the HH technique. The $diag\{\cdot\}$ function returns a matrix with the input vector on the main diagonal and the $triu\{\cdot\}$ function extracts the upper triangular part of an input matrix, excluding main diagonal.

#### C. Givens Rotations

Givens Rotations (GR) can also be employed to factorize the matrix $\mathbf{H}$ [11]. This technique is usually used in embedded systems because of its numerical stability [12]. The general principle of this technique is to cancel the elements of the matrix $\mathbf{H}$ so that a triangular form is obtained. Therefore, GR technique is the only technique which is not iterative and, hence, has a parallelism capabilities that will be explored later on.

Algorithm 3 depicts a MATLAB-like algorithmic description of the GR technique, where the rotation matrix $\Theta$ is calculated such that the elements of $\mathbf{H}$ are cancelled column by column, and an upper triangular matrix $\mathbf{R}$ is finally obtained.

---
[1]Conventional MATLAB notation are employed in the introduced algorithms *i.e.* $\mathbf{A}_{n,n}$ corresponds to the $(n, n)$-th entry of the matrix $\mathbf{A}$, $\mathbf{A}_{:,n}$ corresponds to the $n$-th column-vector of $\mathbf{A}$ and $\mathbf{A}_{n:N,n}$ corresponds to entries from $n$ to $N$ of the $n$-th column-vector of $\mathbf{A}$

**Algorithm 2** Householder triangulation technique.
1: INPUT: $\mathbf{H}$
2: OUTPUT: $\tilde{\mathbf{y}}$, $\mathbf{R}$
3: Initialization: $\mathbf{Q} \leftarrow \mathbf{H}, \mathbf{R} \leftarrow \mathbf{I}_{n_T}$
4: **for** $i = 1, \cdots, n_T$ **do**
5: $\quad [\mathbf{v}, \beta] \leftarrow \mathbf{house}(\mathbf{Q}_{i:n_R, i})$
6: $\quad \mathbf{Q}_{i:n_R, i+1:n_T} \leftarrow \mathbf{Q}_{i:n_R, i+1:n_T} - \mathbf{v}\mathbf{v}^H \mathbf{Q}_{i:n_R, i+1:n_T}$
7: **end for**
8: $\mathbf{R} \leftarrow diag\{\beta\} + triu\{\mathbf{Q}_{1:n_T, 1:n_T}\}$
9: **for** $i = n_T, \cdots, 1$ **do**
10: $\quad \mathbf{Q}_{i:n_R, i:n_T} \leftarrow \mathbf{Q}_{i:n_R, i:n_T} - \mathbf{v}\mathbf{v}^H \mathbf{Q}_{i:n_R, i:n_T}$
11: **end for**
12: $\tilde{\mathbf{y}} \leftarrow \mathbf{Q}^H \mathbf{y}$

**Algorithm 3** Givens rotations algorithm.
1: INPUT: $\mathbf{H}$
2: OUTPUT: $\tilde{\mathbf{y}}$, $\mathbf{R}$
3: Initialization : $\mathbf{Q} \leftarrow \mathbf{H}, \mathbf{R} \leftarrow \mathbf{I}_{n_T}$
4: **for** $i = 1, \cdots, n_T$ **do**
5: $\quad$ **for** $j = i+1, \cdots, n_R$ **do**
6: $\quad\quad \alpha \leftarrow \frac{\mathbf{Q}_{i,\ i}}{\|\mathbf{Q}_{i:j,\ i}\|}$
7: $\quad\quad \beta \leftarrow \frac{\mathbf{Q}_{j,\ i}}{\|\mathbf{Q}_{i:j,\ i}\|}$
8: $\quad\quad \Theta \leftarrow \begin{bmatrix} \alpha & \beta \\ -\beta & \alpha \end{bmatrix}$
9: $\quad\quad [\mathbf{R}_{i,\ i:n_T};\ \mathbf{R}_{j,\ i:n_T}] \leftarrow \Theta[\mathbf{R}_{i,\ i:n_T};\ \mathbf{R}_{j,\ i:n_T}]$
10: $\quad\quad [\mathbf{Q}_{1:n_R,\ i};\ \mathbf{Q}_{j:n_R,\ j}] \leftarrow [\mathbf{Q}_{1:n_R,\ i};\ \mathbf{Q}_{1:n_R,\ j}]\Theta^H$
11: $\quad$ **end for**
12: **end for**
13: $\tilde{\mathbf{y}} \leftarrow \mathbf{Q}^H \mathbf{y}$

### D. Parallel Givens Rotations

The Parallel GR (PGR) technique allows independent operations in the QR decomposition process so that the processing speed is increased. An example of the PGR is explained in the case of a $4 \times 4$ real matrix. The conventional GR technique obtains the following matrix:

$$\mathbf{R} = \begin{bmatrix} r_{1,1} & r_{1,2} & r_{3,1} & r_{4,1} \\ 0_1 & r_{2,2} & r_{3,2} & r_{4,2} \\ 0_2 & 0_4 & r_{3,3} & r_{4,3} \\ 0_3 & 0_5 & 0_6 & r_{4,4} \end{bmatrix}, \quad (6)$$

where the entries $r_{i,j}$ represent the elements of $\mathbf{H}$ that are modified during the QRD. The notation $0_k$ represents the order $k$ in which the corresponding elements of the matrix $\mathbf{H}$ are cancelled out and such that $\{r_{i,j} \mid 1 \leq i,\ j \leq n_T\} \subset \mathbb{C}$ and $\{r_{i,i} \mid 1 \leq i \leq n_T\} \subset \mathbb{R}$. Hence, we remark that 6 sequential and dependent steps are necessary for the cancellation stage for a $4 \times 4$ real matrix.

Equation (7) illustrates the PGR technique to obtain the triangular matrix $\mathbf{R}$ from $\mathbf{H}$:

$$\mathbf{R} = \begin{bmatrix} r_{1,1} & r_{1,2} & r_{3,1} & r_{4,1} \\ 0_1 & r_{2,2} & r_{3,2} & r_{4,2} \\ 0_2 & 0_3 & r_{3,3} & r_{4,3} \\ 0_3 & 0_4 & 0_5 & r_{4,4} \end{bmatrix}, \quad (7)$$

where 2 elements are simultaneously cancelled at the step 3 due to their independence. Hence, a parallelization gain is obtained which increases with the size of the matrix $\mathbf{H}$.

Fig. 1, which consists of 8 pipes, shows the advantages of the PGR. The white blocks represent a free tube while the dark blocks represent a tube being doing the calculation. This calculation may correspond to the calculation of the rotation matrix, to the cancellation of the imaginary part of an element of $\mathbf{R}$, of the real and imaginary parts of an element of $\mathbf{R}$, or the calculation of $\mathbf{Q}$. The potential parallelism has been evoked and can be associated to the QRD complexity reduction.

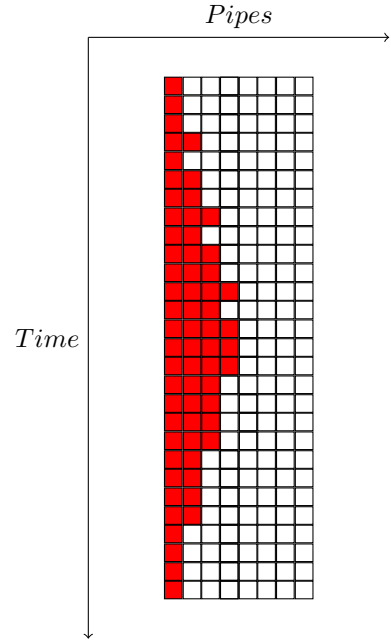

Fig. 1: Parallelism illustration with a QRD process on a $4 \times 4$ matrix made of real entries.

### E. Reduced Complexity of Parallel Givens Rotations

The main idea of the Reduced Complexity PGR (RCPGR) technique is to perform the computational operations in parallel, for both the real an imaginary parts of $\mathbf{H}$, and the matrix $\mathbf{Q}$ is not explicitly obtained.

*Lemma 1:* Consider a modified $n_R \times (n_T + 1)$ input matrix constructed as $[\mathbf{H}\ \mathbf{y}]$ and with $n_R \geq n_T$, the QRD process on the $n_T$ first rows results in a $n_T \times (n_T + 1)$ triangular matrix corresponding to $[\mathbf{R}\ \tilde{\mathbf{y}}]$, with $\tilde{\mathbf{y}} = \mathbf{Q}^H \mathbf{y}$.

*Proof:* Let us consider the initialization step of the QRD algorithm by introducing the $n_R \times (n_T + 1)$ matrix $\underline{\mathbf{Q}} = [\mathbf{H}\ \mathbf{y}]$

**Algorithm 4** Reduced complexity Givens rotations algorithm.

1: INPUT: $\mathbf{H}$
2: OUTPUT: $\tilde{\mathbf{y}}$, $\mathbf{R}$
3: Initialization : $\mathbf{Q} \leftarrow [\mathbf{H}\ \mathbf{y}], \mathbf{R} \leftarrow \mathbf{I}_{n_T+1}$
4: **for** $i = 1, \cdots, n_T + 1$ **do**
5:   **for** $j = i+1, \cdots, n_R$ **do**
6:     $\alpha \leftarrow \frac{\mathbf{Q}_{i,\ i}}{\|\mathbf{Q}_{i:j,\ i}\|}$
7:     $\beta \leftarrow \frac{\mathbf{Q}_{j,\ i}}{\|\mathbf{Q}_{i:j,\ i}\|}$
8:     $\Theta \leftarrow \begin{bmatrix} \alpha & \beta \\ -\beta & \alpha \end{bmatrix}$
9:     $[\mathbf{R}_{i,i:n_T+1}; \mathbf{R}_{j,i:n_T+1}] \leftarrow \Theta[\mathbf{R}_{i,i:n_T+1}; \mathbf{R}_{j,i:n_T+1}]$
10:   **end for**
11: **end for**

and the $(n_T+1) \times (n_T+1)$ matrix $\underline{\mathbf{R}} = \mathbf{I}_{n_T+1}$.
For the sake of simplicity, let us consider the common GS technique. By processing the QRD from columns 1 to $n_T$, the following results are obtained: $\underline{\mathbf{Q}} = [\mathbf{Q}\ \mathbf{y}]$ and $\underline{\mathbf{R}} = \begin{bmatrix} \mathbf{R} & \mathbf{0} \\ \mathbf{0} & \mathbf{1} \end{bmatrix}$. The QRD processing final step is realized at the $n_T + 1$ column. By considering the GS technique, the $\underline{\mathbf{R}}$ coefficients are obtained with the expression $\underline{\mathbf{R}}_{i,\ j} = (\underline{\mathbf{Q}}_{:,\ i})^H \underline{\mathbf{Q}}_{:,\ j}$, $\forall i \in [\![1,\ n_T]\!]$, $\forall j \in [\![1,\ n_T+1]\!]$. Hence $\underline{\mathbf{R}}_{i,\ n_T+1} = (\underline{\mathbf{Q}}_{:,\ i})^H \underline{\mathbf{Q}}_{:,\ n_T+1} = (\underline{\mathbf{Q}}_{:,\ i})^H \mathbf{y}$ at this step of the algorithm process.
It has been previously shown $\underline{\mathbf{Q}}_{:,\ i} = \mathbf{Q}_{:,\ i}$ $\forall i \in [\![1,\ n_T]\!]$. Consequently $\underline{\mathbf{R}}_{i,\ n_T+1} = \mathbf{Q}_{:,\ i}^H \mathbf{y}$ and $\underline{\mathbf{R}}_{1:n_T,\ n_T+1} = [\mathbf{R}\ \tilde{\mathbf{y}}]$. ∎

Lemma 1 implies the inputs of the detection stage, *i.e* $\tilde{\mathbf{y}}$ and $\mathbf{R}$, are obtained without explicitly calculating the unitary matrix $\mathbf{Q}$. This technique is depicted in Algorithm 4 in a MATLAB-like description. This considerably reduces the computational complexity as will be seen in the following section.

## IV. SIMULATION RESULTS

### A. Computational Complexities

Computational complexities[2] of the previously evoked QRD techniques have been theoretically obtained and are displayed in Table I. Herein, we consider the computational complexity to decompose a $n_R \times n_T$ complex matrix.

The corresponding computational complexities are plotted in Figure 2 and the $\tilde{\mathbf{y}}$ received vector computation is also taken into account, which is not the case in general. These simulation results roughly match the classical theoretical complexities [13], but they are more realistic in any Digital Signal Processor (DSP) fashion programming. GR are more complex in terms of the required number of operations compared to STGS: for a $8 \times 8$ complex matrix, the additional complexity is 164%. This difference is due to the need for explicitely computing the $\mathbf{Q}$ matrix. Also, results show that PCPGR achieves a computational complexity reduction of 39% and 77% compared to STGS and GR, respectively, all for a $8 \times 8$ complex matrix.

---
[2]The assumptions are the following: a real product is denoted 1 MUL, a real addition as 0 MUL (Multiply Accumulate (MAC) operation), a real division as 16 MUL (conditional add-subtract algorithm) and a real square root as 32 MUL (dynamic number between $\frac{1}{4}$ and 1 and Taylor series development)

| QRD | MUL |
|---|---|
| STGS | $4n_T^2 n_R + 34 n_R n_T + 32 n_T$ |
| HH | $8 n_R n_T^2 + 6 n_R n_T - \frac{8}{3} n_T^3 + 3 n_T^2 + \frac{395}{3} n_T$ |
| GR | $17 n_R^2 n_T - n_R n_T^2 + 134 n_R n_T - \frac{7}{3} n_T^3 - 73 n_T^2 - \frac{212}{3} n_T$ |
| PGR | $17 n_R^2 n_T - n_R n_T^2 + 134 n_R n_T - \frac{7}{3} n_T^3 - 73 n_T^2 - \frac{212}{3} n_T$ |
| RCPGR | $n_R^2 n_T + 7 n_R n_T^2 + 154 n_R n_T - \frac{7}{3} n_T^3 - 81 n_T^2 - \frac{236}{3} n_T$ |

TABLE I: Complexities equivalences.

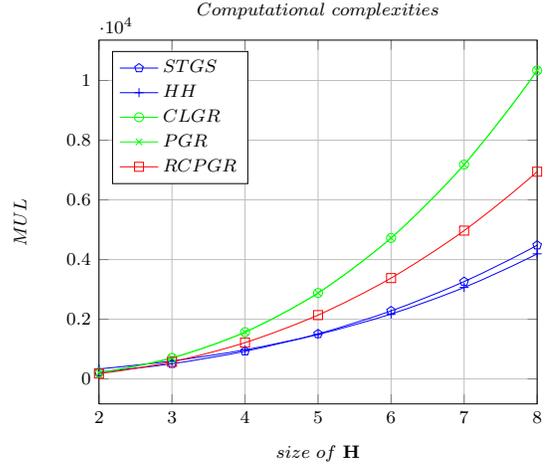

Fig. 2: Computational complexities of the summarized techniques as a function of the size of the to decompose complex matrix.

Results on computational complexities have been presented in the paper and offer strictly the same performance. This point is verified in the Subsection IV-C with the LTE-A parameters background.

### B. Algorithmic Parallelism

Let us define the parallelism gain as the ratio of the number of parallel operations by the total number of operations. The parallelism gain of the summarized QRD algorithms is plotted in Figure 3, with in reference the STGS-based QRD and the proposed RCPGR-based QRD, as a function of the size of the decomposed complex matrix; the size of the channel matrix ranges from $2 \times 2$ to $8 \times 8$, which are the available modes in the LTE-A standard.

The proposed RCPGR-based QRD is more complex, in terms of computational complexity, compared to all the classical QRD. However, thanks to the parallelism potentiality of the algorithm, the proposed technique is the best in terms of execution time, which is an essential point for implementation considerations. Performances are shown to be strictly the same in Section IV-C.

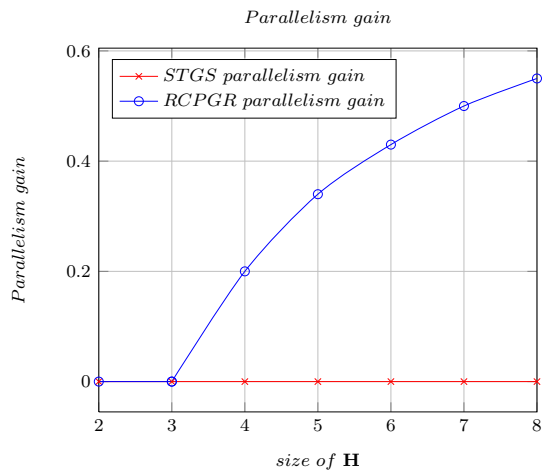

Fig. 3: Algorithmic parallelism as a function of the size of the to decompose complex matrix.

*C. Performances*

Performances results in terms of uncoded BER are given in this part of the paper, for both the previously described SIC and QRD-M techniques. The channel matrix is considered to be perfectly known at the receiver. In Figure 4, the RCPGR QRD-based SIC and QRD-M (M from 2 to 4) performance are plotted as a function of the Signal-to-Noise Ratio (SNR) for a $8 \times 8$ MIMO system, under complex Rayleigh fading channel.

The proposed RCPGR QRD-based SD performance is classically compared to the reference Zero Forcing (ZF) and ML. Note that in terms of error performance, all the introduced QRD techniques lead to similar performance since ordering was not included. The ZF equalizer is considered since it provides the performance upper bound for the lowest possible computational complexity. The ML detector is also considered since it provides performance lower bound, but for an exponential computational complexity. The ML is practically infeasible in the LTE-A background. For example, by considering a QPSK modulation on each transmit antenna and for a $8 \times 8$ system configuration, $4^8 = 65536$ Euclidean distances computations are necessary to decode 16 bits only. The QRD-M performances have been shown to be close to that of the optimum ML, for a polynomial mean complexity and with a parallel preprocessing using the RCPGR algorithm.

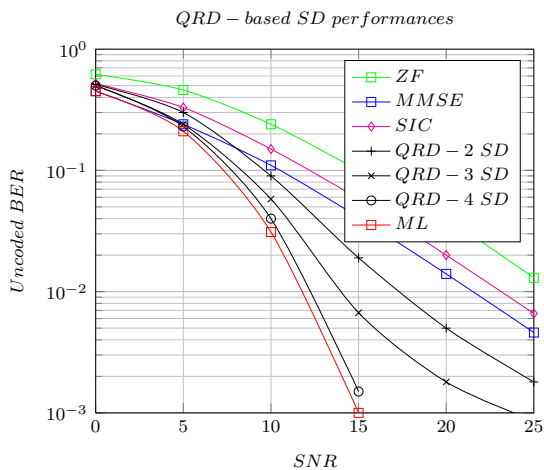

Fig. 4: Uncoded BER of the reference ZF, MMSE, SIC and ML and RCPGR-based QRD-2/3/4 SD as a function of the SNR, for a MIMO $4 \times 4$ Rayleigh channel and QPSK modulations on each layer.

V. CONCLUSIONS

In this paper, several QR decomposition techniques have been addressed. The computational complexities of those algorithms as well as their methodology have been investigated. Our attention in this paper has also been paid to parallelism capabilities of the channel factorization where the parallelization gain of the Givens rotations technique, i.e., RCPGR, has been outlined. The performances of the detection algorithms can be improved by using signal ordering, which was the topic of several studies in the literature.

ACKNOWLEDGMENT

This work was supported by ST-ERICSSON.